\documentclass[11pt]{article}
\pdfoutput=1
\usepackage{amsmath, amsfonts, amssymb}
\usepackage{comment}
\usepackage{graphicx}
\usepackage{psfrag}
\usepackage{amsthm}
\usepackage[usenames,dvipsnames,svgnames,table]{xcolor}
\usepackage{enumerate}
\usepackage{tcolorbox}
\usepackage{mathtools}
\usepackage{soul}
\usepackage{subfigure}
\usepackage{mathrsfs}  
\usepackage{a4wide}
\usepackage{booktabs}
\usepackage{tikz}
\usepackage{tikz-cd}
\usepackage{makecell}
\usepackage{multicol}

\usepackage{color}
\definecolor{dark-gray}{gray}{0.20}
\definecolor{gray}{gray}{0.30}
\definecolor{light-gray}{gray}{0.80}
\definecolor{dark-red}{rgb}{0.7,0,0}
\definecolor{dark-green}{rgb}{0.1,0.4,0}
\definecolor{dark-blue}{rgb}{0.3,0.3,0.7}
\definecolor{light-blue}{rgb}{0.8,0.8,1}
\definecolor{swamp}{RGB}{240, 199, 197}

\usepackage{pifont}
\usepackage{setspace}

\newcommand{\be}{\begin{equation}}
\newcommand{\ee}{\end{equation}}

\def\be{\begin{equation}}
\def\ee{\end{equation}}
\def\bea{\begin{eqnarray}}
\def\eea{\end{eqnarray}}

\def\simleq{\; \raise0.3ex\hbox{$<$\kern-0.75em
		\raise-1.1ex\hbox{$\sim$}}\; }
\def\simgeq{\; \raise0.3ex\hbox{$>$\kern-0.75em
		\raise-1.1ex\hbox{$\sim$}}\; }

\numberwithin{equation}{section}

\usepackage{jheppub} 
\hypersetup{
	colorlinks=true,
	linkcolor=dark-blue,
	citecolor=dark-red,
	urlcolor=dark-green,
	linktoc=page,
	pageanchor=false
}

\title{\centering
Cosmological Phase Transitions and the Swampland
}

\author{Gerben Venken$^1$}

\affiliation{Rudolf Peierls Centre for Theoretical Physics, University of Oxford\\ Beecroft Building, Clarendon Laboratory, Parks Road, OX1 3PU, UK}
\emailAdd{gerben.venken@physics.ox.ac.uk}

\abstract{
I consider the Festina Lente Swampland bound and argue taking thermal effects, as for instance occur during reheating, into account significantly strengthens the implications of this bound. I argue that the confinement scale should be higher than a scale proportional to the vacuum energy, while Festina Lente without thermal effects only bounds the confinement scale to be above the Hubble scale. For Higgsing of nonabelian gauge fields, I find that the magnitude of the Higgs mass should be heavier than a bound proportional to the Electroweak scale (or generally the scale set by the Higgs VEV). The measured values of the Higgs in the SM satisfy the bound. A way to avoid the bound being violated during inflation is to have a large number of species becoming light. If one wants the inflationary scale to lie below the species scale in this case, this bounds the inflationary scale to be $\ll 10^5$ GeV. These bounds have phenomenological implications for BSM physics such as GUTs, suggesting for example a weak or absent gravitational wave signature from the GUT Higgsing phase transition.

}

\begin{document}

\makeatletter
\let\old@fpheader\@fpheader

\makeatother

\maketitle

\section{Introduction}\label{intro}

In recent years much effort has gone into making swampland bounds on low-energy effective theories more phenomenologically and cosmologically relevant. A conjecture which has proven useful in this context is the Festina Lente (FL) bound. This conjecture states that in a quasi-de Sitter vacuum with positive vacuum energy $V$ with a long-range (at the hubble scale) U(1) gauge field with gauge coupling $g$, all charged particles should obey \cite{Montero:2019ekk,Montero:2021otb}
\begin{equation}
\label{eq:FL}
    m^2 > \sqrt{2} g q V^{1/2}\,,
\end{equation}
with $m$ the mass and $q$ the charge of the particles. A series of further papers have explored the phenomenological implications of this bound in various directions \cite{DallAgata:2021nnr,Gonzalo:2021zsp,Lee:2021cor,Cribiori:2022sxf,Ban:2022jgm,Guidetti:2022xct,Montero:2022jrc,Mishra:2022fic,Mohseni:2022ftn,Dalianis:2023ewd,Mohseni:2023ogd}. A point stressed in \cite{Montero:2021otb} is that FL can also be applied to nonabelian gauge theories. For a nonabelian gauge theory, the gauge fields themselves are charged under $U(1)$ subgroups of the nonabelian gauge group. If the gauge interaction is long-range, this then leads to an automatic violation of FL. In the presence of Higgsing or confinement, the nonabelian gauge fields do not lead to a long-range interaction on the Hubble scale and arguments for FL cannot be applied. It then follows that all nonabelian gauge fields in a theory with positive vacuum energy should either be Higgsed or confined, with the Higgsing or confinement scale heavier than the Hubble scale.

We refer to \cite{Montero:2019ekk} for a detailed treatment or \cite{Montero:2021otb,Grana:2021zvf} for a summary of the argument for FL. To briefly sketch the logic of the argument: One considers the evaporation of charged black holes of cosmological size in quasi-de Sitter space, called charged Nariai black holes \cite{1950SRToh..34..160N,Romans:1991nq}. One demands that these black holes evaporate back to empty de Sitter space and do not evolve into geometries with naked singularities. Such singular behaviour would violate cosmic censorship. Further, as empty de Sitter space is the state of maximal entropy, the black hole not evaporating back to it would seem in contradiction with thermodynamics. When Eq. \ref{eq:FL} is violated the radiation produced during the evaporation of charged cosmic sized black holes results in an evolution to a singular geometry rather than empty de Sitter space and from this one obtains the FL bound.

There is more to the cosmic history of our universe than quasi-de Sitter space. After the era of inflation (or some alternative very early cosmic era) and before our own cosmic era, reheating is expected to have taken place, greatly increasing the temperature of the universe and leading to an era where the evolution of the universe was dominated by thermal radiation which gradually cooled down. One then expects that if one has a cosmologically relevant theory of quantum gravity, then if one adds a bath of thermal radiation to the background and evolves forward in time, the resulting cosmic evolution should be consistent with quantum gravity and not violate any swampland bound.

The goal of this short note is to combine Festina Lenta with a background of thermal radiation and derive the resulting swampland constraints, focusing specifically on nonabelian gauge fields. We will see that adding the thermal background will significantly strengthen the implications of FL. The structure of this note is as follows: 

In Sec. \ref{sec:Confinement} we derive our main bounds. We start in Sec. \ref{sec:finiteTFL} by working out under what conditions one can apply FL in the presence of thermal radiation. We then in Sec. \ref{sec:confinement} work out the resulting constraints on confining gauge theories and in Sec. \ref{sec:higgsings} work out the constraints in Higgsed gauged theory. We show that the Standard Model (SM) satiesfies the resulting constrains during the current cosmic era.

In Sec. \ref{sec:Inflation} we explore the implications of our bound during the inflationary era. One way in which our bounds can be satisfied during the inflationary era is by having a large number of particle species become light. In this case if one wishes the inflationary scale to lie below the species scale, it is necessary for inflation to happen at a low scale; below $10^5$ GeV.

In Sec. \ref{sec:caveats} we discuss two caveats to the logic of the preceding sections. In Sec \ref{sec:thermaldestab} we consider the possibility that finite-temperature effects destabilize the vacuum and discuss how this affects our constraints. We consider a string theory example where such destabilization happens and argue that our constraints are automatically satisfied. In Sec. \ref{sec:decoupled} we discuss the possibility of having multiple sectors where the interaction between the sectors is sufficiently weak such that through the expansion of the universe the different sectors do not thermally equilibrate and analyse how this changes our bounds.

We close with the discussion in Sec. \ref{sec:discussion}, summarizing our results, discussing their implications such as how our bounds might constrain BSM physics such as GUTs, discussing how our constraints fit in with the rest of the swampland program and providing an outlook of how our bounds might be applied to concrete models.


\section{Constraining Confinement and Higgsings}
\label{sec:Confinement}

\subsection{Applying Festina Lente at Finite Temperature}
\label{sec:finiteTFL}

Consider adding to a quasi-de Sitter cosmology a bath of thermal radiation at finite temperature $T$, higher than the de Sitter temperature\footnote{This happens for instance during reheating, but we will not assume that our thermal bath is at the reheating temperature, we simply add radiation at some temperature $T$ as initial conditions as a thought experiment and demand that subsequent evolution is consistent with the swampland conjectures.}. Thermal effects will contribute effective terms to the potential. At sufficiently high temperature, these thermal effects will make it so that the Higgs potential has a minimum at $\phi=0$ where electroweak symmetry is restored. Similarly, at high temperatures QCD confinement is lost. Throughout this text, we will often refer to a theory which is Higgsed at low energy as the `electroweak theory' and a theory which is confined at low energy as `QCD' and speak of the electroweak/QCD scale. Except where we explicitly consider real-world values, we merely use these terms as a convenient way to refer to generic theories with Higgsing/confinement.

In the introduction we explained that FL forbids such electroweak symmetry restoration as it renders the $U(1)$ charge carriers massless. One might then be tempted to conclude that FL forbids temperatures sufficiently high to deconfine / to restore symmetry. However, this conclusion is premature. The argument for FL relies on the evaporation of black holes in a quasi-de Sitter background. Introducing thermal radiation can alter the cosmic evolution. To ensure that we are allowed to apply FL, we should ensure that we perform our analysis in a quasi-de Sitter phase. let us check what this implies.

The energy density of thermal radiation in $d$ spacetime dimensions is given by
\begin{equation}
    \rho_R = \sigma T^d\,,
\end{equation}
where the prefactor $\sigma$ is defined as a numerical coefficient, $\pi^2 /15$ for $d=4$, times the number of degrees of freedom which are thermally excited. We will take $d \geq 4$ such that charged Nariai black holes exist, but otherwise we leave $d$ arbitrary for the moment, although we are especially interested in the case $d=4$.

The energy density of the vacuum energy is given by
\begin{equation}
    \rho_V = V\,,
\end{equation}
where the potential $V$ is taken to be either stabilized at a positive CC or sufficiently slowly rolling for FL to apply.

In order for FL to apply we must then demand that the vacuum energy is one of the driving components of cosmological evolution such that the cosmological Nariai black holes necessary for the FL argument exist. This implies that for FL to apply one must have
\begin{equation}
\label{eq:Tbound}
    \frac{\rho_R}{\rho_V} \leq c \implies T \leq \left(\frac{c V}{\sigma}\right)^{1/d}\,.
\end{equation}
Here $c$ is an unknown, possibly model-dependent, order one parameter. When $\rho_R \ll \rho_V$, it is clear that the vacuum energy dominates cosmic evolution, while for $\rho_V \ll \rho_R$ it is clear that radiation dominates the cosmic evolution. For $\rho_R \approx \rho_V$ the situation is murkier; at which precise ratio of $\rho_R$ and $\rho_V$ the charged Nariai black holes necessary for the FL argument do (not) exist will depend on the parameters, such as gauge couplings, of the model one considers. We introduce $c$ to account for this ambiguity.

Note that (for $\sigma \sim \mathcal{O}(1)$) the temperature bound Eq. \ref{eq:Tbound} is significantly higher than the de Sitter temperature $T_{dS}\sim H$.

If the bound Eq. \ref{eq:Tbound} is obeyed, it is possible to treat the cosmology as quasi-de Sitter and one may consider quasi-Nariai black holes evaporating in the presence of a background of thermal radiation.

We then wish to apply the FL bound based on the evaporation of quasi-Nariai BHs in the presence of this background of thermal radiation. There are two asides we must make here to legitimize this:

First, due to cosmic expansion the thermal radiation dilutes and cools during the BH evaporation process. One should then in principle treat the radiation temperature as time-dependent in the BH evaporation analysis. It was shown in \cite{Montero:2019ekk} that unless the $U(1)$ gauge coupling is extremely small, $g \lesssim H/M_p$, Nariai BHs discharge in less than one e-folding. So long as timescale is shorter than an e-folding, the radiation temperature remains of the same order of magnitude and one may treat the radiation temperature as effectively constant, which we will do.

Second, the presence of background radiation will affect the Nariai black hole decay process. Will this alter our conclusions? Recall how the charged Nariai BH discharges in the rapid regime: The electric field between the BH and cosmic horizon is rapidly converted into neutral radiation. This neutral radiation evolves to produce a big crunch singularity. FL follows from demanding that this big crunch does not occur. Adding additional thermal radiation as a background to this analysis will only help the big crunch to form. The conclusion of FL is then unchanged and if anything one needs to make the FL bound even stronger in the presence of this additional thermal radiation. (Assuming as said before that the radiation energy one adds is small enough such that the spacetime is quasi-dS initially and one can talk of Nariai BHs)

We then conclude that when the temperature of the background radiation obeys \eqref{eq:Tbound}, it is legitimate to apply the FL bound. In particular, there should not be a vacuum where EW symmetry is restored or QCD is deconfined at these temperatures.

\subsection{Confinement}
\label{sec:confinement}

Let us first consider the case of confinement. Let $\Lambda_{QCD}$ be the confinement energy scale. When $T>\Lambda_{QCD}$, the theory deconfines. In order not to violate FL, the deconfinement temperature must be above the temperature bound Eq. \ref{eq:Tbound}. This impies
\begin{equation}
\label{eq:confinementbound}
    \Lambda_{QCD} > \left(\frac{c V}{\sigma}\right)^{1/d}\,.
\end{equation}
If $c$ and $\sigma$ are taken to be $\mathcal{O}(1)$ constants, this implies that the confiment scale should be above the vacuum energy scale. This holds for real-world QCD and is a significantly stronger constraint than the zero-temperature FL constraint of \cite{Montero:2021otb} which demanded $\Lambda_{QCD} > H$.

One may also consider the case where $\sigma$ is not assumed to be $\mathcal{O}(1)$. One has $\sigma \sim N_T$, with $N_T$ the number of species lighter than $T$ such that they are part of the thermal radiation. One has $N_{T} \leq N_{sp}$, with $N_{sp}$ the number of species of particles light enough to be in the EFT. The species scale \cite{Dvali:2007hz,Dvali:2009ks,Dvali:2010vm,Dvali:2012uq} is given by
\begin{equation}
\label{eq:species}
    \Lambda_{sp} \sim \frac{M_{p} }{ N_{sp}^{1/(d-2)}} \,.
\end{equation} 
and this energy scale should be the UV cut-off scale of an EFT with $N_{sp}$ species. Combining $\left(c V \sigma\right)^{1/d} > H$ and $\sigma \sim N_T \lesssim N_{sp}$ one finds $c^{1/(d-2)}\Lambda_{sp}\gtrsim H$, such that Eq. \ref{eq:confinementbound} is a stronger constraint than $\Lambda_{QCD} > H$ whenever the species scale is at higher energies than the Hubble scale, which is necessary for the EFT to be controlled as the Hubble scale sets the cosmological horizon scale and serves as an IR cut-off.

\subsection{Higgsing}
\label{sec:higgsings}

In this section we will restrict for simplicity to $d=4$. Consider a zero-temperature potential for the Higgs field $\phi$ of the form
\begin{equation}
    V(\phi) = - \mu^2 |\phi|^2 + \lambda |\phi|^4 + \frac{\mu^4}{4 \lambda}\,,
\end{equation}
where $\mu^2, \lambda >0$ and the constant term is such that the EW-breaking vacuum has zero vacuum energy\footnote{This is legitimate when as in the real world the hilltop is at an energy many orders higher than the EW breaking vacuum}. The potential at the top of the hill at zero temperature is then
\begin{equation}
    V(0) = \frac{\mu^4}{4 \lambda}\,.
\end{equation}
At zero temperature the mass of the Higgs at $\phi=0$ is negative, $-\mu^2$, and one has a tachyonic instability. At finite temperature, thermal contributions to the action will contribute an effective thermal to the Higgs mass at $\phi=0$, such that at sufficiently high temperatures the symmetry-restoring vacuum vacuum becomes stable\footnote{Or possibly metastable for a first-order phase transition. For a first-order phase transition, there is a temperature $T_C$ above which the $\phi=0$ vacuum is metastable and the symmetry-breaking vacuum at a lower energy. Above a temperature $T_*>T_C$, the $\phi=0$ vacuum is either the lowest-energy or only vacuum, see e.g. \cite{Mukhanov:2005sc} for a review. For temperatures $T_C<T<T_*$, in order to apply FL it is necessary to make sure the discharge time of the cosmological black holes $t \sim (g^2 V(0))^{-1/4}$ is shorter than the tunneling time to the symmetry-breaking vacuum, while for $T > T_*$ there is no such constraint. As $T_C$ and $T_*$ are generically of the same order of magnitude and our arguments are at the level of parametrics, we will not worry about distinguishing between these and assume there is no issue in applying FL on the basis of tunneling to the symmetry-reabking vacuum.} \cite{Kirzhnits:1972ut,Dolan:1973qd,Weinberg:1974hy}. The temperature at which the mass of the Higgs turns turns from negative to positive due to thermal effects is 
\begin{equation}
    \alpha T_c^2 =  \mu^2\,.
\end{equation}
Here, $\alpha$ is a parameter dependent on the couplings. For instance \cite{Dolan:1973qd}, for a $U(1)$ gauge field $A^\mu$ coupling to the Higgs in such away that the Lagrangian contains amongst other terms an effective mass term for the gauge field $g'^2 \phi^2 A^\mu A_\mu$, there will be a term in $\alpha \sim g'^2$, while the Higgs quartic self-coupling will provide terms $\alpha \sim \lambda$ . For a more complicated theory such as the SM analogous terms appear in $\alpha$ for the other couplings.

One must now demand that at $T_C$ the temperature is high enough that the FL bound does not apply, as otherwise there will be a stable symmetry-restoring vacuum contradicting FL. from Eq. \ref{eq:Tbound} this implies
\begin{equation}
\label{eq:higgsbound1}
     \frac{\mu^2}{\alpha} > \left(\frac{c V(0)}{\sigma}\right)^{1/2} = \sqrt{\frac{c}{4 \lambda \sigma}} \mu^2 \implies \lambda > \frac{c \alpha^2}{4 \sigma}
\end{equation}
In this model, the absolute value of the mass of the Higgs is of order $|m_H|\sim \mu$ both in the symmetric vacuum and in the symmetry breaking vacuum, so we will refer to the scale of both of these masses as $|m_H|$. One sees that our bound implies
\begin{equation}
    \frac{V^{1/4}(0)}{|m_H|} \approx \frac{1}{\lambda^{1/4}} < \left( \frac{4 \sigma }{c \alpha^2} \right)^{1/4}\,.
\end{equation}
The VEV of the Higgs in the EW-breaking vacuum is $ \langle \phi \rangle = \mu / \sqrt{2 \lambda} $ and this sets the electroweak energy scale $E_{EW}$, the mass scale of the Higgsed gauge fields. One then sees
\begin{equation}
\label{eq:higgsscale}
    \frac{E_{EW}}{|m_H|} \approx \frac{1}{\lambda^{1/2}} < \left( \frac{4 \sigma }{c \alpha^2} \right)^{1/2}\,.
\end{equation}
One sees that Festina Lente requires a Higgs mass which is sufficiently heavy compared to the electroweak scale and the symmetry-restoring vacuum energy scale. If $c$, $\sigma$, and $\alpha$ are $\mathcal{O}(1)$, then $|m_H|$ must be at least the same energy scale as $E_{EW}$ and $V^{1/4}(0)$. In the SM, these quantities are all at the same energy scale. At the level of the parametric arguments in this note, we cannot make a sharper statement. it would be very interesting to in the future compute the model-dependent quantities $c$, $\sigma$, and $\alpha$ precisely to obtain a sharp bound on the Higgs mass from Eq. \ref{eq:higgsscale}.

By using $\alpha = \mu^2 / T_C^2$ one may recast Eq. \ref{eq:higgsbound1} as
\begin{equation}
    \frac{T_C}{V^{1/4}(0)} > \left(\frac{c}{\sigma}\right)^{1/4}\,.
\end{equation}
Using the values $T_C \approx 160$ GeV, $V^{1/4}(0) \approx 104$ GeV, derived from the measured couplings assuming a quartic Higgs potential \cite{ParticleDataGroup:2022pth,DOnofrio:2015gop} and estimating $1/\sigma^{1/4} \approx 0.5$, one finds that for the measured values in the standard model the bound gives $3.08>c^{1/4}$. The expectation is that $c$ will be at most $\mathcal{O}(1)$ such that the standard model with measured values obeys our bound.

It then seems that swampland constraints (specifically FL) prefer a heavy Higgs mass over a light Higgs (relative to the electroweak scale). A heavy Higgs leads to a higher-order electroweak phase transition, while a light Higgs leads to a first order electroweak phase transition. From our parametric analysis, we cannot make a definitive statement that first-order Higgs symmetry-breaking phase transitions are in the swampland (the transition between first- and higher-order is in order of magnitude around the scale $E_{EW}$), however our results suggest a preference for a higher-order phase transition.

In the preceding we have focused on the standard quartic Higgs model. The logic of our argument goes through analogously for generic scalar potentials where nonabelian gauge symmetries are restored at a point $\phi=0$. One has a (meta)stable symmetry-restoring vacuum at finite temperature maximum if
\begin{equation}
    V''(0) + \alpha T^2 > 0\,,
\end{equation}
with $V$ the zero-temperature potential. Demanding that the symmetry-restoring vacuum is unstable for all temperatures where FL applies via Eq. \ref{eq:Tbound} implies
\begin{equation}
\label{eq:higssderivbound}
    \frac{-V''(0)}{\sqrt{V(0)}} > \alpha \sqrt{\frac{c}{\sigma}}\,,
\end{equation}
where we assume $V(0)>0$. Note that so long as $\alpha / \sqrt{\sigma} > H / M_P$, this bound implies that the refined de Sitter conjecture \cite{Garg:2018reu,Ooguri:2018wrx} holds at all points in the scalar moduli space where a nonabelian gauge symmetry is restored. Alternatively, if one assumes the refined de Sitter conjecture at $\phi=0$, this bounds the couplings (via $\alpha$) and the number of species lighter than $T$ (via $\sigma$).
\section{Constraining Inflation}
\label{sec:Inflation}

In Sec. \ref{sec:confinement} we saw that, for real-world values, QCD has a characteristic energy scale above the vacuum energy scale and handily satisfies the constraint Eq. \ref{eq:confinementbound}. However, if the universe went through an early period of inflation, the inflationary vacuum energy scale $V_{inf}$ could have been significantly higher than the QCD scale.

In order to satisfy finite-temperature FL for the confining theory we must have
\begin{equation}
\label{eq:QCDinfbound}
    \Lambda_{QCD} > \left(\frac{c V_{inf}}{\sigma}\right)^{1/d}\,.
\end{equation}
One drastic way to satisfy this constraint would be to demand $V_{inf}< \Lambda_{QCD}$. However, there is another way: one can increase $\sigma$ by having additional states becoming light. Let $\sigma \sim N_{T}$, with $N_{T}$ the number off states with mass $m<T$ such that they make up part of the thermal radiation. To satisfy FL we must then have
\begin{equation}
\label{eq:infbound1}
      c\left(\frac{ \Lambda_{inf}}{\Lambda_{QCD}}\right)^d \lesssim N_T \,,
\end{equation}
where $\Lambda_{inf} = V_{inf}^{1/d}$ is the inflationary energy scale. It is then clear that in order for inflation to occur at energy scales hierarchically larger than $\Lambda_{QCD}$, a large number of states must become light. As one increases the ratio $\frac{ \Lambda_{inf}}{\Lambda_{QCD}}$ the number of light states grows. The appearance of towers of light states as one moves in parameters space is a familiar feature of the Swampland Program, most notably in the form of the distance conjecture \cite{Ooguri:2006in} when moving in moduli space; which has been generalized to parameters other than moduli such as the cosmological constant in AdS \cite{Lust:2019zwm}. It is worth stressing that the appearance of light states here is a consequence only of considering the evaporation of charged black holes in de Sitter space (FL) and cosmological thermal effects. We have note had to refer to string theory, extra dimensions, or other swampland conjectures to discover the need for a large number of light states. This is a common feature of the web of swampland conjectures where one finds that one conjecture often implies features of other conjectures in specific set-ups.

With a large number of light states $N_T$ in hand, we can start applying other conjectures to the light states. In particular, according to the species scale, Eq. \ref{eq:species},
At temperatures below the EFT cut-off, one must have $N_T \leq N_{sp}$. Using this and combining Eq. \ref{eq:infbound1} with Eq. \ref{eq:species} one finds
\begin{equation}
    \Lambda_{sp}\lesssim c^{-1/(d-2)}\left(\frac{\Lambda_{QCD}}{\Lambda_{inf}}\right)^{d/(d-2)}M_P\,.
\end{equation}
If one wants to be able to describe inflation within the EFT under consideration, one must have $\Lambda_{inf}\ll \Lambda_{sp}$ leading parametrically to (after dropping $c$)
\begin{equation}
    \Lambda_{inf} \ll \Lambda_{QCD}^{d/(2d-2)}M_P^{(d-2)/(2d-2)}\,.
\end{equation}
For $d=4$ and real-world values of the QCD\footnote{We use the current value of $\Lambda_{QCD}$. The inflaton could directly couple to the $\text{Tr}[F\wedge \ast F]$ term of QCD, changing the coupling strength of QCD during inflation. This will change the bound given here. As an extreme case, one could propose raising $\Lambda_{QCD}$ high enough in this tower to satisfy Eq. \ref{eq:QCDinfbound} directly without invoking a large number of light states as an alternative way to satisfy FL during inflation.} and Planck scales this leads to $\Lambda_{\inf} \ll 10^5$ GeV, a low scale for inflation. In fact, depending on how large one believes the hierarchy between the inflationary and species scale should be for EFT, the maximal inflationary scale is not very far above the electroweak scale. Such an extremely low inflationary scale has attractive features \cite{German:2001tz} e.g. when it comes to dealing with the moduli problem.

In the preceding we have presented our discussion in terms of the confinement scale, since for real-world values this provides a stronger constraint than working with the electroweak theory. However, the analysis goes through entirely analogously in the case of Higgsing. Start by substituting $\sigma \sim N_T$ in Eq. \ref{eq:higssderivbound},
\begin{equation}
    \frac{- V''}{\sqrt{V_{inf}}} > \alpha \sqrt{\frac{c}{N_T}} > \sqrt{\frac{c}{N_{sp}}} = \alpha \sqrt{c} \frac{\Lambda_{sp}}{M_P}\,.
\end{equation}
Demanding $\Lambda_{inf} \ll \Lambda_{sp}$ one finds
\begin{equation}
    \Lambda_{inf} \ll \left(\frac{-V'' M_P}{\alpha \sqrt{c}}\right)^{1/3} = \frac{|m_H|^{2/3} M_P^{1/3}}{\alpha^{1/3} c^{1/6}}\,.
\end{equation}

Stepping away from the nonabelian gauge fields, one may also ask how matter charged under a $U(1)$ gauge field constrains inflation via Eq. \ref{eq:FL}. Considering the electron with mass $m_e$ and charge $e$ one finds
\begin{equation}
    \Lambda_{inf} < m_e / (2^{1/4} \sqrt{e})\,.
\end{equation}
Doesn't this imply that the inflationary scale is bound to be at most ten times the electron mass scale? In fact, this point was already addressed in \cite{Montero:2019ekk}. This constraint on the inflationary scale can be satisfied by having the masses of charged particles or the electromagnetic gauge coupling change during the inflationary era, e.g. by considering a correctly tuned model of Higgs inflation \cite{Rubio:2018ogq}. Nevertheless, it is clear that it will also be easier to satisfy the bound from EM-charged matter if inflation happens at a low scale. Constraints of the ordinary FL bound on inflation were also further explored in \cite{Lee:2021cor}.
\section{Two caveats}
\label{sec:caveats}

\subsection{Thermal vacuum destabilization}
\label{sec:thermaldestab}

An important caveat to the discussion in this paper is that we assume that thermal effects do not destabilize the vacuum. If the vacuum destabilizes at a temperature below/at the scale of deconfinement/Higgsing, then any discussion of physics in the vacuum at temperatures above the destabilization temperature is meaningless.

In string theory there is a well-controlled set-up of confinement in the presence of a positive vacuum energy provides a nice example of such thermal destabilization.

Consider the set-up of Klebanov-Strassler \cite{Klebanov:2000nc,Klebanov:2000hb}. From a five-dimensional gravitational perspective one has an infinitely long throat geometry with a smooth tip. This theory has a holographic dual which is a four-dimensional field theory. The gauge theory in four dimensions is confining, with the confinement holographically resulting from the smoothness of the tip of the throat geometry.

One may now introduce an anti-D3 brane to the tip of the throat as proposed by Kachru, Pearson and Verlinde (KPV) \cite{Kachru:2002gs}. This results in a metastable vacuum with a positive vacuum energy (for an approriate number of branes and fluxes in the throat). One may now consider the KPV set-up at finite temperature. This has been the subject of a series of analyses \cite{Michel:2014lva,Polchinski:2015bea,Bena:2014jaa,Cohen-Maldonado:2015ssa,Bena:2016fqp,Cohen-Maldonado:2016cjh,Armas:2018rsy,Armas:2019asf,Blaback:2019ucp,Nguyen:2019syc,Nguyen:2021srl}. The main take-away for our purposes is, as recently summarized in the review of \cite{VanRiet:2023pnx}, that at sufficiently high temperatures one has a ``donut-berliner" transition. In this transition the geometry at the tip of the throat becomes singular and confinement is lost. However, at the same time the vacuum destabilizes. In this case, FL is automatically obeyed for any temperature where one can speak of a metastable vacuum, providing a stringy example where the finite-temperature version of FL holds.

\subsection{Many species, thermal equilibrium, and Decoupled Sectors}
\label{sec:decoupled}

In the preceding sections we have repeatedly considered the regime where a large number of particle species increases $\sigma$, thereby lowering the maximum temperature at which one may apply FL via Eq. \ref{eq:Tbound}. The possibility of considering this regime played a crucial role in our considerations of inflation. One might object; while we \textit{could} consider thermal radiation where all species lighter than $T$ are part of the radiation, could we not also consider thermally exciting only one particle species, and from there run our arguments using Eq. \ref{eq:Tbound} with only $\sigma$ set to the value where only one species is excited. If that were allowed, it would provide stronger constraints than the argument where all light species are part of the thermal radiation and would invalidate our arguments involving many species of light particle.

The counterargument would be that the particle species all interact and via these interactions the one species that we first excite would start exciting other particle species and we can only speak of equilibrium thermal radiation once all the particle species are in their thermal distribution. This logic holds in flat space. However, in an expanding universe one must have
\begin{equation}
    \Gamma > H\,,
\end{equation}
where $\Gamma$ is the interaction rate for two particle types to exchange energy $T$, in order for two particle species to thermally equilibrate with each other.

If at a given temperature two sectors are sufficiently decoupled, they do not need to equilibrate. For a theory with $S$ decoupled particle sectors, one may then run the logic of the preceding sections for radiation from each of these sectors separately\footnote{Just as reheating served as the physical process motivating why one should consider the presence of such thermal radiation in preceding sectors, we may here imagine a rolling scalar coupled to only one sector as leading to the production of thermal radiation in just that sector.}. For each sector $i=1,..,S$ one then separately uses Eq. \ref{eq:Tbound} with a distinct $\sigma_i$. one obtains the strongest constraints by considering the smallest $\sigma_i$, that is the sector with the smallest number of particle species. 

However, if the confined/Higgsed nonabelian gauge field to which one wishes to apply FL is in a different decoupled sector than the thermal radiation, then the gauge field will not feel a thermal background and will not become deconfined/unhiggsed due to thermal corrections. 

One concludes that for a gauge field in decoupled sector $i$ one should then in the preceding sections substitute $\sigma$ with the $\sigma_i$ of that sector. As a result, if for instance we want inflation not to violate FL by having a large number of species as proposed in Sec. \ref{sec:Inflation}, then it is not just necessary to have a large number of total species: each decoupled sector with nonabelian gauge fields witch characteristic energy scales below the inflationary scale should separately have an appropriately large number of species.
\section{Discussion}
\label{sec:discussion}

In this note I have explored the consequences of applying Festina Lente in a finite temperature background (higher than the de Sitter temperature). As with the ordinary FL bound, nonabelian gauge fields in a (quais-)de Sitter background are required to be either confined or Higgsed at low energies. However, the addition of a bath of thermal radiation to FL significantly strengthens the constraints one can derive.

For confinement, we have seen that the confiment energy scale must be above an energy scale proportional to the vacuum energy, Eq. \ref{eq:confinementbound}, which is a significantly stronger result than the bound from just FL, which only required the confinement scale to be above the Hubble scale. The implications of $\Lambda_{QCD}>H$ were explored in \cite{Mishra:2022fic,Mohseni:2023ogd}, it would be interesting to further explore the consequences of Eq. \ref{eq:confinementbound} analogously.

For Higgsing, we have seen that the magnitude of the Higgs mass $|\mu |$ in the symmetry-restoring vacuum is required to be heavy in terms of a scale proportional to the Higgs VEV $\langle \phi \rangle$, Eq. \ref{eq:higgsscale} as well as heavy compared to a scale proportional to the symmetry-restoring vacuum energy scale, Eq. \ref{eq:higssderivbound}. We saw that the measured values of the Standard Model Higgs satisfy this bound. Our bound should also apply to Higgsing in a GUT or hidden sector and will constrain BSM physics. In particular first-order Higgsing phase transitions are at risk of violating our bound, although we cannot fully determine whether first-order phase transitions are completely excluded by our bound without determining the $\mathcal{O}(1)$ model-dependent coefficients involved. First-order GUT phase phase transitions lead to interesting physics, for instance as a catalyst of baryogenesis or as a source of a gravitational wave signature (see e.g. \cite{Weir:2017wfa,Caprini:2018mtu,Mazumdar:2018dfl,Croon:2019kpe} for recent reviews). The visibility of such a gravitational wave signature grows with increased $\langle \phi \rangle / |\mu|$, while our bound implies that this ratio cannot be too large, suggesting a weak or absent gravitational wave signature. In the future it would be interesting to work out more precisely what our bound implies for the phenomenology of specific  GUT models.

For a small number of light particle species the FL bound seems likely to be violated during inflation. A violation of FL during inflation can be avoided by having a large number of species becoming light, Eq. \ref{eq:infbound1}. In this case, if one wants the inflationary energy scale to lie below the species scale, the inflationary scale needs to be below $10^5$ GeV. It has previously been stressed in the context of the swampland program that during inflation one is at risk of a large number of light states, appearing for instance via the distance conjecture via the large field ranges involved in inflation, see e.g. \cite{Ooguri:2006in,Agrawal:2018own,Hebecker:2018vxz,Bedroya:2019snp}. In particular, \cite{Bedroya:2019tba} argued for an upper bound on the scale of inflation of $10^9$ GeV based on restricting field ranges, with which our preference for low-scale inflation is in line. It is remarkable that we have also encountered the appearance of a large number of light states without directly using these conjectures, string theory, or extra dimensions as an input.

 The inflationary scale should be below the species scale in order to describe both our current cosmological era and the inflationary era within the same EFT. This does not necessarily imply that the inflationary scale has to be below the species scale. An alternative would be for the inflationary era and the current cosmological era to be described by distinct EFTs. When in one EFT breaks down it could be that after integrating out the states above the species scale the other EFT arises as a controlled description, along the lines of the emergence proposal \cite{Ooguri:2018wrx,Heidenreich:2017sim,Harlow:2015lma,Heidenreich:2018kpg,Grimm:2018ohb}. When the effective degrees of freedom are altered so drastically between the early and late universe, it may be that the early-universe effective theory does not resemble traditional inflation. Alternatively, one could, as recently investigated in \cite{Agrawal:2020xek}, have an early topological phase such as in string gas cosmology \cite{Brandenberger:1988aj}.

In Sec. \ref{sec:thermaldestab} we have considered the possibility that thermal effects destabilize the vacuum. If the vacuum is destabilized at temperatures before the deconfinement / symmetry-restoring temperature then this ensures that our thermal version of FL is automatically satisfied. In a string theory context, we saw that the KPV set-up provides an example where this happens. It would be interesting to further study our constraints in the context of concrete string compactifications. For instance, one issue in stringy inflation is the overshoot problem \cite{Brustein:1992nk} and a low inflationary scale (in line with our constraints) may help to resolve the overshoot problem. On the other hand, \cite{Conlon:2022pnx} has recently argued that in the context of LVS compactifications \cite{Balasubramanian:2005zx,Conlon:2005ki} on the contrary a large hierarchy between the inflationary and electroweak scale may be useful in resolving the overshoot problem.

To sharpen the constraints which have appeared in this note, an obvious step would be to compute the order one parameter $c$ in Eq. \ref{eq:Tbound}. This parameter is set by the ratio between the energy in thermal radiation and the vacuum energy at which charged black holes of size the cosmological horizon still exist. As a crude first estimate, one might suppose that for Nariai black holes to exist, the FLRW metric which should serve as a background for the black hole should have a cosmological horizon. This requires an effective equation of state for radiation combined with vacuum energy $w<-1/3$. Working out the resulting constraint on $c$ on finds $c<1$. To find the exact value of $c$ in a specific model, it is necessary to construct the corresponding Nariai-like black hole geometry with radiation in that concrete model. It would be especially interesting to work this out for the standard model as this would allow us to turn our bound on the Higgs mass into a concrete numerical lower bound.

\section*{Acknowledgements}
I thank Arthur Hebecker for valuable feedback on an early draft. I thank Arthur Hebecker, Miguel Montero, Georges Obied, and Thomas Van Riet for valuable discussion. This work was supported by funding from an STFC consolidated grant, grant reference ST/X000761/1.

\bibliographystyle{JHEP}
\bibliography{refs}

\end{document}